\documentclass[conference]{IEEEtran}
\IEEEoverridecommandlockouts
\usepackage{cite}
\usepackage{amsmath,amssymb,amsfonts}
\usepackage{algorithmic}
\usepackage{graphicx}
\usepackage{textcomp}
\usepackage{xcolor}
\usepackage{subcaption}

\def\BibTeX{{\rm B\kern-.05em{\sc i\kern-.025em b}\kern-.08em
    T\kern-.1667em\lower.7ex\hbox{E}\kern-.125emX}}

\usepackage[strict]{changepage}

\usepackage{framed}

\definecolor{formalshade}{rgb}{0.95,0.95,1}

\begin{document}

\title{Exploration of LLMs, EEG and behavioral data to measure and support attention and sleep\\
}

\author{\IEEEauthorblockN{Akane Sano}
\IEEEauthorblockA{ 
\textit{Rice University}\\
USA \\
akane.sano@rice.edu}
\and
\IEEEauthorblockN{Judith Amores}
\IEEEauthorblockA{\textit{Microsoft Research}\\
USA \\
judithamores@microsoft.com}
\and
\IEEEauthorblockN{Mary Czerwinski}
\IEEEauthorblockA{\textit{Microsoft Research}\\
USA \\
marycz1031@gmail.com}
}

\maketitle

\begin{abstract}
We explore the application of large language models (LLMs), pre-trained models with massive textual data for detecting and improving attention and sleep. We investigate the use of LLMs to estimate attention states, sleep stages, and sleep quality and generate sleep improvement suggestions and adaptive guided imagery scripts based on electroencephalogram (EEG) and physical activity data (e.g., waveforms, power spectrogram images, numerical features). Our results show that LLMs can estimate sleep quality based on human textual behavioral features and provide personalized sleep improvement suggestions and guided imagery scripts; however, detecting attention, sleep stages, and sleep quality based on EEG and activity data requires further training data and domain-specific knowledge.

\end{abstract}


\section{Introduction}

Human altered states such as attention and sleep play significant roles in health \cite{buysse2014sleep}, safety \cite{horne1995sleep}
, and productivity\cite{rosekind2010cost}
. By precisely measuring these states, we can design adaptive tools and interfaces that respond effectively to users and help promote their health. 

Human attention states have been measured using physiological and behavioral data such as electroencephalogram (EEG) \cite{richer2018real}, facial expressions \cite{baltruvsaitis2011real}, and eye tracking\cite{vidal2012wearable}. 
. Measuring human attention states can help design systems that enhance driver alertness, minimize interruptions during focus, or promote relaxation before sleep.

Extensive research has explored computational methods for measuring, evaluating, and improving sleep. For example, many algorithms have been developed to estimate sleep quality and stages using human physiological and behavioral sensor data including EEG and motion\cite{imtiaz2021systematic}. Computational systems have been designed to promote better sleep\cite{shinMobilePhoneInterventions2017}.

Recent advances in natural language processing have leveraged massive textual data to train large language models (LLMs). %Researchers have explored LLMs for diverse tasks including detection, generation, and question-answering\cite{}.
Some studies have used LLMs for understanding human physiological and behavioral data and designing health applications including EEG abnormality detection and wearable sensor-based sleep quality detection\cite{kim2024eeg}\cite{kim2024health,huExploringLargeScaleLanguage2024}. 

LLMs hold promise for health applications including human altered state detection and personalized feedback delivery; however, rigorous evaluations have not been conducted, particularly regarding the integration of different human physiological and behavioral data (e.g., waveforms, numerical features, power spectrogram images) into LLMs for understanding the potential, accuracy, limitations, and reliability of the models.

In this paper, we evaluate LLMs for detecting and supporting human attention and sleep. Our ultimate goal is to create personalized, adaptive systems that enhance individuals' attention and sleep, similar to multisensory frameworks like \cite{amoresAImageryMultisensoryApproach2024}. To achieve this, in this paper, we conduct early explorations by integrating biobehavioral data into LLMs to understand their capabilities. We ask the following two research questions in the paper: 

\begin{enumerate}
    \item Can LLMs interpret/sense attentive states, sleep stages, and sleep quality?
    \item Can LLMs provide personalized and adaptive feedback to help improve sleep?
\end{enumerate}

We investigate the impact of various time scales and different input modalities of EEG, motion, and textual data on LLMs' performance, reasoning, and generated responses for detecting attention states, sleep stages, and sleep quality detection and improving sleep.

This work provides the first systematic evaluation of LLMs across multiple biobehavioral modalities (EEG waveforms, spectrograms, numerical features, actograms, and self‑report text) for attention and sleep applications. We identify modality‑specific limitations and outline design principles for future multimodal LLM‑based health systems.

%\begin{figure}
%    \centering
%    \includegraphics{test.pdf}
%    \caption{if space is available, will show a diagram to feed data into LLM and get output}
%    \label{fig:enter-label}
%\end{figure}

\section{Methods}

We describe experiments and datasets for 1) user state detection and 2) sleep improvement suggestion generation to address RQ 1 and 2.

\subsection{Experiment 1: user state detection}
We conduct three different detection tasks, a) attention detection, b) sleep stage detection, and c) sleep quality detection.

\subsubsection{Datasets \& Data processing}
We use the following datasets for our experiments.

\begin{figure}

     \begin{subfigure}[b]{0.45\linewidth}
         \centering
         \includegraphics[width=\linewidth]{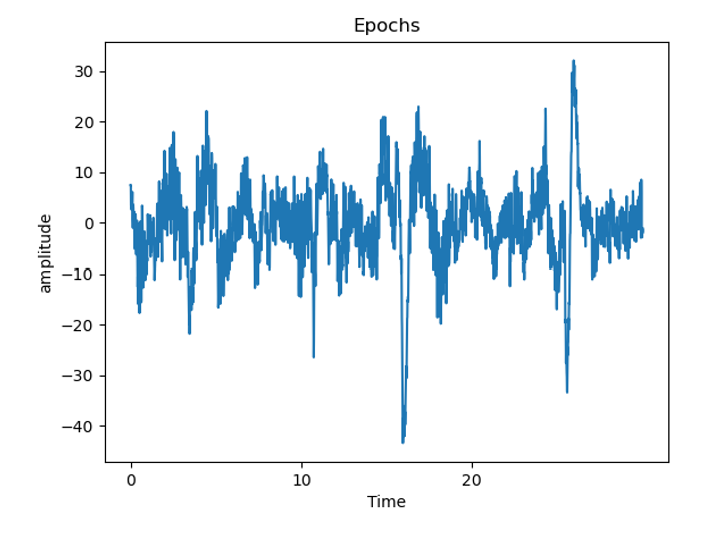}
         \caption{EEG Waveform }
         \label{fig:EEGwave}
     \end{subfigure}
\hfill
     \begin{subfigure}[b]{0.45\linewidth}
         \centering
         \includegraphics[width=\linewidth]{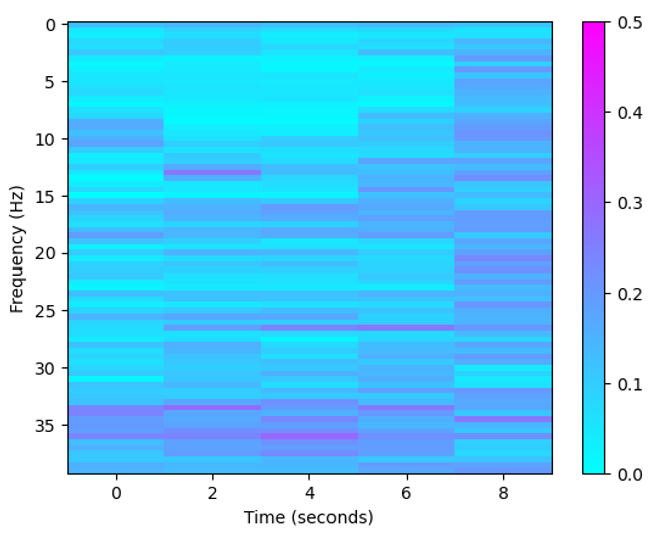}
         \caption{EEG Spectrogram}
         \label{fig:EEGpower}
     \end{subfigure}

         \begin{subfigure}[b]{0.45\linewidth}
         \centering
         \includegraphics[width=\linewidth]{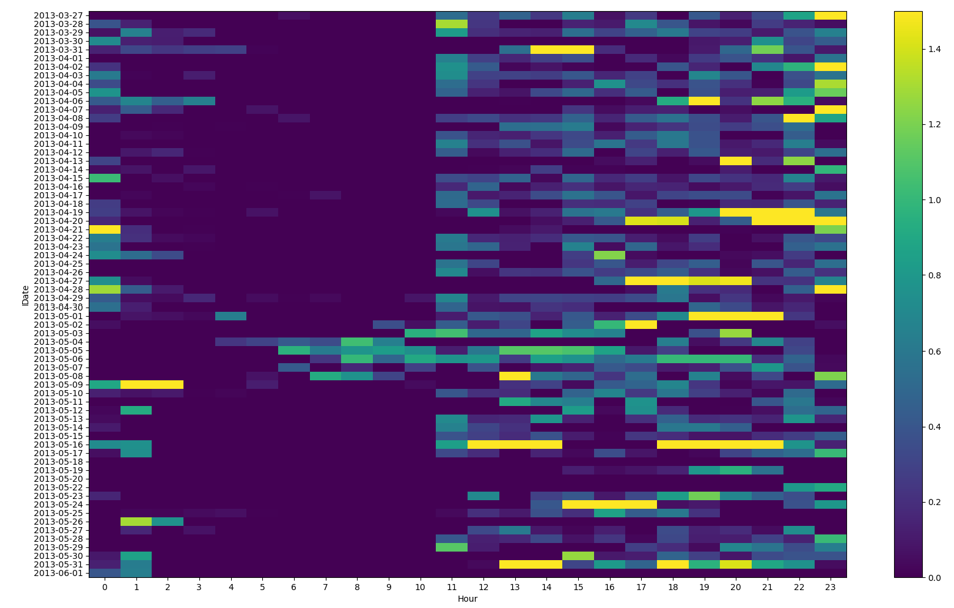}
         \caption{Actogram (x axis: time 0am-0am), y-axis: dates, z-axis:physical activity levels)}
         \label{fig:actograms}
     \end{subfigure}
\hfill
           \begin{subfigure}[b]{0.45\linewidth}
         \centering
         \includegraphics[width=\linewidth]{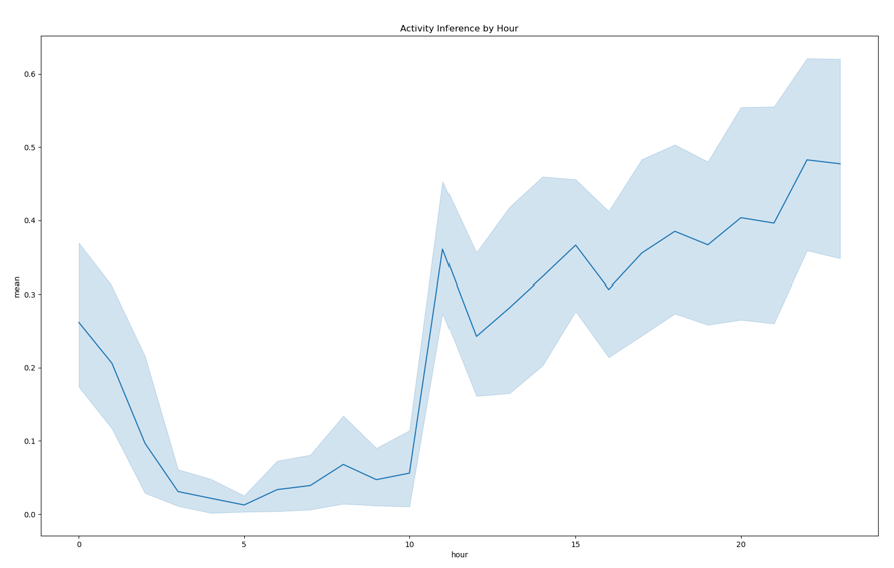}
         \caption{Averaged activity graph (x axis: time 0am-0am), y-axis: physical activity levels)}
         \label{fig:averaged}
     \end{subfigure}
    % \label{fig:imagedata}
\caption{Image Data Fed into LLMs}
\end{figure}

\textbf{Mental Attention State\cite{aci2019distinguishing}:}
This dataset contains 25 hours of EEG data collected using 14-channel Emotive. Five participants were engaged in a low-intensity task of controlling a computer-simulated train. Three mental states were observed in this study: focused, unfocused, and drowsy. We merged unfocused and drowsy into an unfocused state. 
We prepared three different types of information for attention detection (focused vs unfocused/drowsy): a) filtered EEG data (Fig. \ref{fig:EEGwave}): Raw EEG signals were processed using a bandpass Butterworth filter (order: 128, 0-40Hz), b) time-frequency spectrograms(Fig. \ref{fig:EEGpower}): These were computed using wavelet transform (Daubechies wavelet, every 10 sec) to provide frequency components over time, and c) 11 features: these features include power spectrum density (delta, theta, alpha, beta), amplitude, standard deviation, kurtosis, alpha/delta, theta/alpha, delta/theta, the 90th percentile amplitude.
We segmented the data into 10-sec intervals, resulting in 919 training samples, 230 validation samples, and 287 test samples.

\textbf{Sleep EDF expanded\cite{867928}:}
This dataset contains 197 nights of polysomnography data collected from individuals aged 18-101 years. The data include EEG from two channels, Fpz-Cz and Pz-Oz, EOG, EMG, and event markers. Sleep stages are labeled as follows: 0 (Wake), 1 (stage 1), 2 (stage 2), 3 (stages 3 \& 4), and 4 (REM sleep).
We segmented the data into 30-sec epochs (training set: 152362 epochs, Validation set: 38092 epochs, test set: 500 epochs). We used Fpz-Cz for sleep stage detection. We used the same input types as those used for mental attention states: a) filtered EEG data, b) time-frequency spectrograms, and c) 11 features.

\textbf{Student Life\cite{wang2014studentlife}:}
This dataset contains mobile phone sensor and survey data collected from 46 college students. We used the Pittsburgh Sleep Quality Index (PSQI) (19 self-rated questions including sleep-related behaviors and self-reported sleep quality) administered both at the pre and post study and its scoring rules to categorize each participant as a poor or good sleeper. We use physical activity data collected from participants' phones to compute participants' daily activity levels and patterns. We assess sleep quality detection (good vs poor) using the following inputs: a) participants' textual responses to PSQI questions, b) physical activity-based actograms: visual representations of 24 hour activity levels over days(Fig. \ref{fig:actograms}), and c) physical activity-based hourly averaged graphs: hourly levels and variations in physical activity (Fig. \ref{fig:averaged}).

%\begin{figure}
%\label{fig:EEGwavepower}
%     \centering
%     \begin{subfigure}[t]{0.6\linewidth}
%         \centering
%         \includegraphics[width=\linewidth]{fig/actogram.png}
%         \caption{Actogram (x axis: time 0am-0am), y-axis: dates)}
%         \label{fig:EEGwave}
%     \end{subfigure}
%\hfill
%     \begin{subfigure}[t]{0.49\linewidth}
%         \centering
%         \includegraphics[width=\linewidth]{fig/activitygraph.png}
%         \caption{Averaged activity graph}
%         \label{fig:EEGpower}
%     \end{subfigure}

%\end{figure}

\subsubsection{Models}
We compare various LLMs and traditional machine learning models. LLMs:
1) Zero-shot learning: we feed data (EEG waveform images, spectrograms, or features) into LLMs. No specific training is conducted and the LLMs leverage their pre-existing knowledge. We use two LLM variants: GPT 4 vision (2024-02-15-preview) for image input and GPT 4 (2023-05-15) for textual input, 2) In-context learning LLM: we include input data and label examples in prompts so that LLMs (GPT 4) learn from context and adapt their response accordingly. 3) Fine-tuned LLM: We finetune LLMs (GPT 3.5 turbo 2024-02-15-preview) using training and validating datasets. 
A traditional machine learning model, %1) Neural network (2 hidden layers, 128 neurons in each layer), 2) 
XGBoost uses a technique called gradient boosting that combines simple decision trees for accurate predictions. We also analyze feature importance by looking at the number of times each feature is used for trees.
3) Baseline (majority vote): simply predicts the majority class for all test samples. It serves as a basic reference point to evaluate the performance of other models. 
For sleep quality detection, we also use the ground truth scoring method based on PSQI. We evaluate model performance using accuracy and weighted F1 score. Please see examples of LLM prompts in the Appendix.

\subsection{Experiment 2: Personalized and adaptive sleep improvement feedback}
We explore whether LLMs can generate personalized content for sleep improvement and focus on generating 1) sleep improvement suggestions and 2) guided imagery scripts. To generate the sleep suggestions, we feed LLMs various user context or profile information including a) EEG features from Sleep EDF dataset, b) PSQI answers in Student Life dataset, c) physical activity-based actograms, d) gender, e) age group, f) ethnicity, g) health issues such as “anxiety”, “PTSD”, “insomnia”, “pain”, “awakenings during nights”, “shift worker”, h) user preferences such as favorite environments: beach, forest, favorite animals: dogs, cats; hobbies: traveling, baseball, exercise. To generate scripts to help a person sleep, we prompt the LLM to leverage a Guided Imagery technique \cite{utay2006guided}. This technique is also known as Guided Visualization, a mind-body technique that involves visualizing positive images or scenarios using all senses to help promote relaxation, manage anxiety, and stress, and enhance well-being. A typical session might include relaxation exercises, quiet sitting or lying down, and focusing on a specific goal. See prompt examples in the Appendix.

\section{Results}

\subsection{User state detection}
\subsubsection{Attention detection}

Table \ref{tab:attention} summarizes attention detection performance. Traditional machine learning models outperformed LLM-based models for attention detection. Among the LLM models, fine-tuned GPT-3.5 models demonstrated the best performance. 
However, the GPT-4 vision model faced limitations. It failed to infer user states in approximately 25-30\% of cases, often returning a generic response such as ``As a text-based AI, I do not have the capability to process images and I cannot assist with this request". In addition, when we feed high dimensional EEG features, the GPT model indicated that it requires machine learning models and did not infer user states. 

The fine-tuned GPT3.5 model typically shows high confidence levels of 100\%. The model shows information about the power spectrum and ratios (e.g., delta, theta, alpha, beta, alpha/delta) as the explanations for inference.
For example, when correct inference is made with 100\% confidence, the GPT 3.5 model explains ``The participant is unfocused because the alpha/delta ratio is high and the beta power is low, which are indicative of a relaxed and unfocused state."

Incorrect inferences lack clear explanations and appear based on simple heuristic rules.
For example, when ground truth is focused but unfocused is inferred, the model said ``The participant’s state is unfocused because the alpha/delta ratio for channel AF3 is 0.166027, which is below the threshold for focused state."
When ground truth is unfocused but focused is inferred, the model said ``The participant is focused because the alpha/delta ratio is high and the beta power is low, which are characteristic of a focused state."

%"The alpha and beta power in the AF3 channel are relatively high compared to the delta and theta power, which is characteristic of an unfocused state".

We also analyze feature importance from the XGBoost model and learn that the model uses more variety of features for classifying focus vs unfocus including channel O2 kurtosis, AF4 beta, F3 percent 90, F7 beta, and T8 standard deviation.

%correct inference: [high confidence; 100%]
%The participant is unfocused because the alpha/delta ratio is high and the beta power is low, which are indicative of a relaxed and unfocused state."

%Incorrect inference:
%[GT: focused, inference: unfocused] 
%The participant’s state is unfocused because the alpha/delta ratio for channel AF3 is 0.166027, which is below the threshold for focused state.
%[GT: unfocused, inference: focused] 
%The participant is focused because the alpha/delta ratio is high and the beta power is low, which are characteristic of a focused state.

\begin{table}[]
\caption{Attention Detection}
\begin{tabular}{|l|l|l|l|}
\hline
Models                   & acc     & F1   & note\\ \hline
Finetuned GPT3.5 (features)        & 53.7    &  0.51 &\\ \hline
In-context GPT4v (spectrogram)       & 55.7  & 0.56   & 25\% ``cannot assist"   \\ \hline
In-context GPT4 (features)       & 50.0 &  0.57  & 97\% ``cannot assist"   \\ \hline
Zero-shot GPT4v (spectrogram)                & 66.2  &   0.55   & 30\%  ``cannot assist"\\ \hline
Zero-shot GPT4 (features)                & -  &   -   & 99\% ``cannot assist"\\ \hline
%Neural networks          &  &  &\\ \hline
XGBoost (features)           & 91.3  & 0.93  %need update
& \\ \hline
Baseline (majority vote) & 66.6  &  0.53&\\ \hline
\end{tabular}
\label{tab:attention}
\end{table}

\subsubsection{Sleep stage detection}
\begin{figure}
\includegraphics[width=\linewidth]{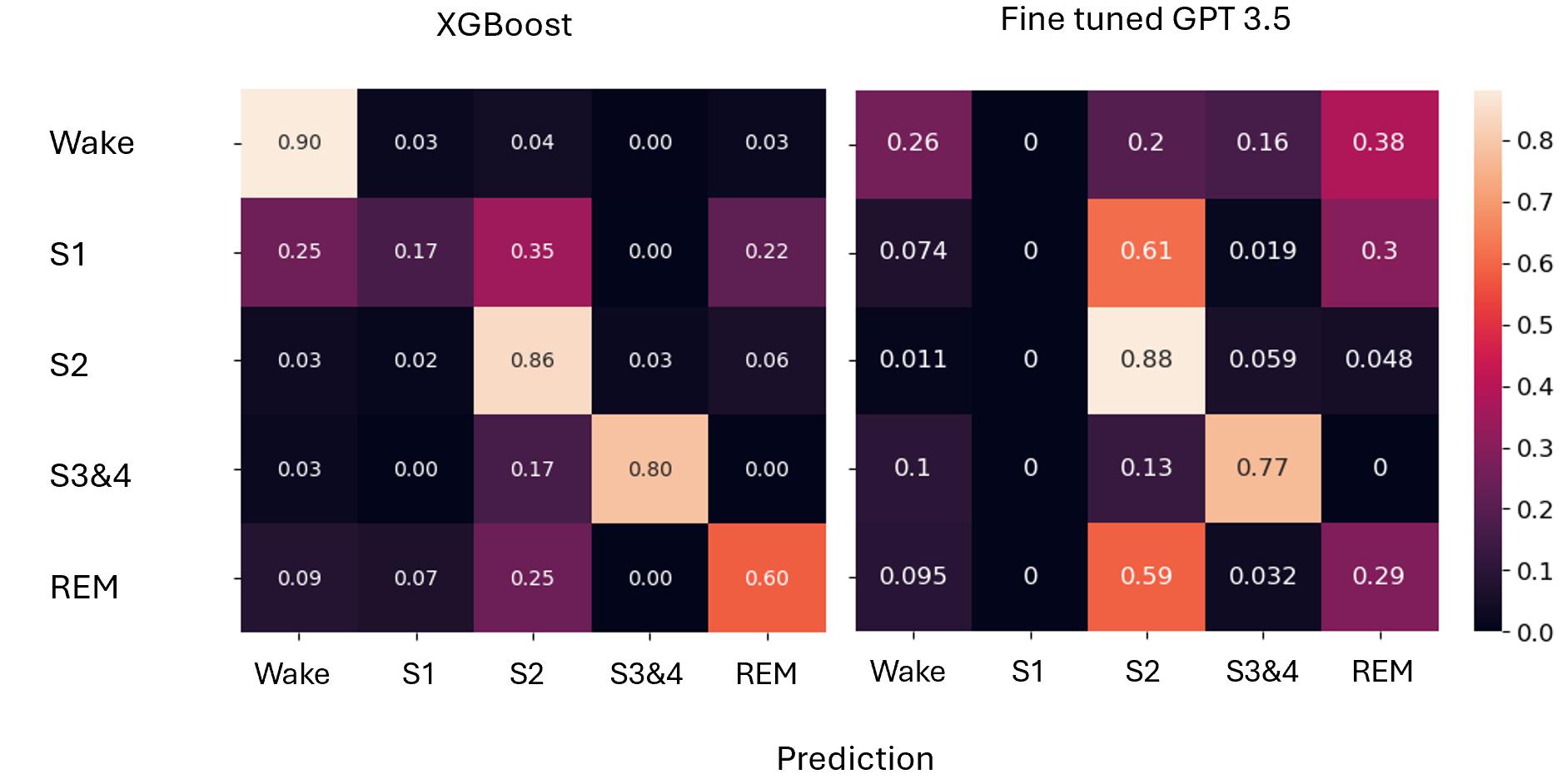}
    \caption{Sleep stage detection models: confusion matrix}
    \label{fig:sleepcm}
\end{figure}

Table \ref{tab:sleepstage} summarizes sleep stage detection performance.
LLM-based sleep detection shows lower performance than ML-based models. 
GPT 4 vision models (both in-context learning and zero-shot learning) often return a message that they cannot assist with the task. In-context learning with EEG waveforms performs worse than zero-shot learning with EEG waveforms. In-context learning with EEG features seems more effective than the zero-shot LLM model with EEG features. GPT fine-tuned models show higher misclassification rates compared to XGBoost. For example, they misclassify wake as REM, stage 1 as stage 2, and REM as stage 2. However, stage 2 and stage 3 \& 4  are similar (Fig.\ref{fig:sleepcm}).

The fine-tuned GPT 3.5 model shows inference confidence from 20 to 100 \% (mean: 79.7 \%, median 80\%, stdev: 9.7\%). The model outputs the explanation that it uses power spectrum and ratio for classifying sleep stages.
For example, when the model infers correct stage 3\&4 sleep, it outputs ``The dominant frequency bands in the EEG signal are delta and theta, with high power in both bands. The alpha and beta power are relatively low. The alpha delta ratio is low, indicating a higher proportion of slow wave activity. The theta alpha ratio is high, suggesting a predominance of theta activity. The delta theta ratio is also high, indicating a higher proportion of slow wave activity compared to theta activity. These features are indicative of stage 3 or 4 sleep, also known as slow wave sleep."

The important features extracted from the XGBoost model show kurtosis, mean, standard deviation, and 90th percentile amplitude in addition to power spectrum and ratio features.

%power spectrum, ratio, kurtosis, standard deviation, mean, and 90th percentile of amplitude

\begin{table}[]
\caption{Sleep Stage Detection}
\begin{tabular}{|l|l|l|l|}
\hline
Models                   & acc     & F1  & note\\ \hline
Finetuned GPT3.5 (features)       & 50.6    & 0.46 & \\ \hline
In-context GPT4v (EEG signals)        & 25.4  & 0.21  & 51\% ``cannot assist"  \\ \hline
In-context GPT4 (features)      & 44.7  &  0.40  &  8\% ``cannot assist"  \\ \hline
Zero-shot GPT4v (EEG signals)             & 19.2  & 0.19  &  13\% ``cannot assist"   \\ \hline
Zero-shot GPT4  (features)             & 29.0  & 0.25  &  0.2\% ``cannot assist"   \\ \hline
%Neural networks          & 76.31 & 0.74 &\\ \hline
XGBoost  (features)            & 75.6  &  0.74  &  \\ \hline
Baseline (majority vote)  & 37.2  & 0.20 &\\ \hline
\end{tabular}
\label{tab:sleepstage}
\end{table}

%[high confidence]
%80\% The dominant frequency bands in the EEG signal are delta and theta, with high power in both bands. The alpha and beta power are relatively low. The alpha delta ratio is low, indicating a higher proportion of slow wave activity. The theta alpha ratio is high, suggesting a predominance of theta activity. The delta theta ratio is also high, indicating a higher proportion of slow wave activity compared to theta activity. These features are indicative of stage 3 or 4 sleep, also known as slow wave sleep.

%incorrect inference
%[GT: wake] 
%The features indicate a high power in the alpha, theta, beta, and gamma frequency bands, as well as a high alpha delta ratio and theta alpha ratio. These patterns are consistent with REM sleep, which is characterized by high brain activity and vivid dreaming. The confidence level is 80%.

\subsubsection{Sleep quality detection}
We compare sleep quality detection (good vs poor sleep; confidence levels) using two approaches: (1) scored by PSQI standard scoring procedure vs. 2) GPT4 (Table \ref{tab:sleepquality}). Overall accuracy of the GPT4 model is 85.7\%. 
There are 11 error cases out of 77. GPT-4 demonstrates relatively good sleep quality detection performance based on textual answers to PSQI; however, GPT-4 struggles to recognize sleep quality accurately when participants show poor sleep behavior (e.g., short sleep, difficulty falling asleep, or disrupted sleep) but good self-reported sleep quality (inference confidence is also low (60-65 \%). GPT seems to emphasize subjective sleep quality over answers to other PSQI questions.
Confidence levels of the GPT fine-tuned model are 60-100 \% (mean: 80\% and std: 7.9 \%).
The confidence level is high when self-reported sleep quality is consistent with sleep behaviors (e.g., very bad self-reported sleep quality and poor sleep related behaviors such as long sleep latency, awakenings during the night, bad dreams, pain, feeling too cold, worry, and difficulty maintaining enthusiasm for daily activities).

%6 hrs sleep, minor sleep issues, self-report sleep quality “good”
%One case: 4-5 hr sleep, minor sleep issues, self-report sleep quality “good”

LLMs detect sleep and active periods and regular sleep patterns from the physical activity actograms and averaged graphs provided; however, the image data fed into LLMs are recognized as poor sleep. LLMs easily pick up some physical activities during the night time that occurred only a few nights out of 60 days of data and recognize them as a sign of poor sleep. For instance, when the actograms show increased activity after 5 am for a few days out of 60 days of data, suggesting potential wakefulness, LLMs misinterpret this as poor sleep quality.
This implies GPT-4's limitations about the challenge of understanding a full range of human sleep behavior and variabilities. Sleep quality is multifaceted and LLMs might not fully grasp the nuances.

\begin{table}[]
\caption{Sleep Quality Detection}
\centering
\begin{tabular}{|l|l|l|}
\hline
PSQI scoring                   & GPT4 response
     & \# of participants  \\ \hline
good        & good    & 39 \\ \hline
poor       & poor  & 27  \\ \hline
good & poor  & 0   \\ \hline
poor            & good  &  11 \\ \hline

\end{tabular}
\label{tab:sleepquality}
\end{table}

%Errors: 11 /77 cases

\subsection{Sleep Improvement Feedback Generation}

LLMs modify suggestions and guided imagery scripts based on user input and profiles (e.g., symptoms, user preferences). 
Tables \ref{tab:guidedimagery} and \ref{tab:sleepimprove} show examples of sleep improvement suggestions and guided imagery scripts based on user profiles.

LLMs generate sleep improvement suggestions that align with cognitive behavioral therapy for insomnia (CBT-i)\cite{hofmann2012efficacy}:  
(1) consistent sleep schedule, (2) avoid late night activities, (3) avoid fluids close to bedtime, (4) try relaxation techniques to help fall asleep and improve sleep quality,  (5) restful room environment (light, temperature),
(6) reduce caffeine intake, large meals, alcohol before bed,
and (7) regular exercise. 
LLMs also weave user profiles (e.g., what they like, and issues they have) into generated guided imagery scripts and change scenes and phrases.

When we feed EEG features for generating guided imagery scripts, if we just simply ask LLM to generate a script based on EEG features, LLM incorporates numerical information into guided imagery scripts such as "As you sink deeper into this state of relaxation, there is a harmonious balance between theta and alpha waves with a ratio of 2.3. " For users, hearing specific numerical information about EEG features might not be useful because they might not know how those numbers are associated with their states, therefore, we adjust prompts not to include numerical information in the generated guided imagery script (Table \ref{tab:guidedimagery}, Appendix for a prompt example). Further tuning of prompt design and evaluation are required to generate effective adaptive feedback.
Sleep improvement suggestions do not change based on EEG features because suggestions are not based on momentary EEG features but habitual sleep related behaviors.

\begin{table*}[]
\caption{GPT4-generated sleep improvement suggestions}
\label{tab:sleepimprove}
\begin{tabular}{|l|l|}
\hline
Profiles & LLM-generated Suggestions                                                                                                                                                                                                                                                                                                                                                                                                                                                                                                                                                                                                                                                          \\ \hline \hline
Headaches disturb sleep
           & \begin{tabular}[c]{@{}l@{}}Check with a medical professional about the headaches to see if they can be treated

\end{tabular} \\ \hline
Trying to solve coding problems in bed 

           & \begin{tabular}[c]{@{}l@{}}Avoid activities such as coding which could generate stress and disrupt sleep

\end{tabular}                                                                                           \\ \hline

Bed feels uncomfortable
& \begin{tabular}[c]{@{}l@{}}Maybe consider investing in a new mattress or bedding if your current one feels uncomfortable 

\end{tabular}                                                                                                                            \\ \hline

Stress
& \begin{tabular}[c]{@{}l@{}}Engagement in stress-relieving activities such as yoga, mindfulness,\\ and deep-breathing can help in reducing stress levels which can improve sleep quality
 
\end{tabular}   
\\ \hline

OCD
& \begin{tabular}[c]{@{}l@{}}Consultation with a Sleep Specialist or mental health professional may also be beneficial,\\ especially since OCD was mentioned as a reason for troubled sleep

\end{tabular}   

%\\ \hline
%Age group, gender

%& \begin{tabular}[c]{@{}l@{}}No difference

%; 
%\end{tabular}   
\\ \hline

\end{tabular}

\end{table*}

\begin{table*}[]
\caption{GPT4-generated guided imagery scripts}
\label{tab:guidedimagery}
\begin{tabular}{|l|l|}
\hline

Profiles & LLM-generated Guided Imagery Scripts                                                                                                                                                                                                                                                                                                                                                                                                                                                                                                                                                                                                                                                           \\ \hline \hline
Like Beach           & \begin{tabular}[c]{@{}l@{}}Close your eyes and imagine yourself on a peaceful beach; Begin with lying down in a comfortable position. \\Close your eyes and take a couple of deep breaths, inhaling and exhaling gradually. \\Visualize the calming view of a beach, the sand beneath your feet, the tranquil waves that are rhythmically kissing the shore.\\ Feel the warmth of the sun on your skin, the softness of the sand as it molds perfectly to your feet. \\Hear the distant sound of seagulls, the gentle lull of the waves. Breathe in the refreshing, salty sea air. \\With every breath you take, your body feels lighter and sleepier. \\Let this calming beach imagery help to lower your heart rate and help you drift off to sleep. \\
\end{tabular} \\ \hline
Like Family
           & \begin{tabular}[c]{@{}l@{}}Close your eyes and imagine you're sitting in a cozy, comfortable family room. The room is warm, inviting, and filled with\\ soft, dimmed light that creates a soothing atmosphere; This scene prompts feelings of relaxation, comfort, and safety, associated \\with being surrounded by loved ones.  You see all your loved ones gathered, chatting, and enjoying each other's company. \\There's a gentle hum of relaxed conversation and soft laughter,  the sounds of your family's voices proving to be the most \\calming sounds you've ever heard; Hearing familiar voices has been shown to reduce stress and anxiety, allowing your\\ mind to unwind and merge into a relaxation state suitable for sleep. You feel an overwhelming sense of love, joy, and peace \\as you sit there surrounded by the people who matter the most. Outside the room, you see through the window that it's\\ nighttime with a sky full of stars and the moon shining brightly, creating a perfect scenery indicating it's time for you to sleep....
\end{tabular}                                                                                           \\ \hline

%Forest& \begin{tabular}[c]{@{}l@{}}Now, imagine yourself standing at the entrance of a calm, peaceful forest.\\ The sun is just rising and the light filters through the leaves, casting beautiful, dappled shadows on the forest floor; 
%\end{tabular}                                                                                                                            \\ \hline

PTSD& \begin{tabular}[c]{@{}l@{}} Imagine you're in a beautiful, peaceful forest; Begin by closing your eyes and taking several deep breaths, in through the \\nose and out through the mouth. As you continue these slow, deep breaths, begin to visualize yourself standing at the edge \\of a beautiful, calm forest....This forest is your safe space, a haven untouched by past trauma. You can stay here as long as\\ you need, resting in the peaceful ambiance. No harm can find you here. Continue to cultivate this sense of peace and safety\\ with each deep, calming breath.... 
\end{tabular}   
\\ \hline

Insomnia& \begin{tabular}[c]{@{}l@{}}Imagine yourself in a serene, beautiful forest; As insomnia has been bothering you, positive visualization can aid in releasing\\ negative energy and reduce stress....
\end{tabular}   

\\ \hline
Shift workers
& \begin{tabular}[c]{@{}l@{}}Relaxing Beach Journey; Begin by making yourself comfortable. Slowly take a deep breath in and as you exhale, close your \\eyes. .....For a shift worker, building a regular sleep schedule can be tough. Guided imagery can aid in achieving a relaxed\\ state quicker once in bed, making the most of the sleep time available...
\end{tabular}   

\\ \hline
EEG features
& \begin{tabular}[c]{@{}l@{}}You are now entering the first stage of sleep, a stage of light sleep. This slow transition into sleep is marked by the gentle\\ wave-like rhythm of your alpha and theta powers. Your consciousness begins to dissolve into a calm and cloud-like tranquility...

\end{tabular}   
\\ \hline

\end{tabular}

\end{table*}

\section{Discussion}

\subsection{Summary of findings}
This work explores the usage of LLMs and physiological and behavioral sensor data for attention and sleep detection and sleep improvement. Our experiments highlight both LLMs' strengths and limitations.

LLM-based attention and sleep detection exhibit lower performance compared to traditional ML models. Fine-tuned LLMs improve models' ability to handle diverse contexts. However, we also found limitations. The fine-tuned GPT3.5 model uses limited features (e.g., power spectrum density and ratio) for classification, even after fine-tuning, and GPT-4 vision models fail to handle visual input. Also, feeding high dimensional numerical features such as EEG features to LLMs does not reliably estimate user states. To improve human state detection, LLM's knowledge needs to be extended beyond simple one-on-one relationships such as increased alpha band activity during unfocused state and LLMs require further refinement to handle diverse human physiological and behavioral data, variabilities, and patterns effectively. This might be possible using fine-tuning with larger datasets and retrieving external sources of knowledge such as retrieval-augmented generation (RAG).
In addition, textual knowledge might not fully capture complex physiological and behavioral patterns; therefore, integrating textual information with numerical and visual data is essential in order to understand variability within and across individuals. While LLM models have improved in multimodal reasoning and fine-tuning techniques, challenges remain in effectively integrating diverse data types. Current models still struggle with handling high-dimensional physiological and behavioral data, and fine-tuning alone may not be sufficient for robust generalization. Expanding training datasets and leveraging multimodal RAG approaches could enhance their ability to process complex variability across individuals.

LLM-based sleep improvement suggestions and guided imagery scripts are personalized and adaptive to user profiles. Automatically generated suggestions and scripts have a potential for AI-based conversational systems or intervention systems after effectiveness and safety are carefully tested.

\subsection{Limitations}
There are several limitations in this study. First, this study is 
an early exploration with limited datasets and limited LLMs. Refining
prompts and using newer models and large and diverse datasets might help enhance task performance. 
We intentionally use interpretable features rather than low dimensional embeddings to test the extent to which LLM internal knowledge contributes to altered state detection and improvement. Our study only relies on public datasets and offline experiments; therefore, user studies with end users and clinicians are necessary to evaluate generative responses in terms of accuracy, effectiveness, and safety. Lastly, beyond using or fine-tuning pre-trained LLMs, other advanced approaches include training transformer or other models on large-scale unlabeled and labeled EEG data \cite{cuiNeuroGPTFoundationModel2024,lloydStressGPTStressDetection2024}, leveraging specialized architectures, and employing multimodal learning to enhance user state detection.

\subsection{Ethical Considerations}
Finally, We also discuss the ethical considerations of using LLMs to detect and improve human-altered states.
First, feeding personal physiological and behavioral data to LLMs could raise privacy concerns. Users might worry about the security and confidentiality of their sensitive information. Transparent consent processes and reliable data anonymization are important. Second, LLMs are trained on massive data that might contain biases. Bias detection and mitigation strategies are necessary to ensure fair outcomes. LLMs might have the capability to generate unethical, harmful, or inaccurate content or manipulate individuals. Implementing guidelines for responsible use and monitoring LLM-generated content is required.

\bibliographystyle{IEEEtran}
\bibliography{references}

\section*{Appendix}
%\let\normalsize\footnotesize
%\footnotesize

%\sectionfont{ \footnotesize}
\subsection*{Prompt and LLM response examples}
\subsubsection*{Zero-shot learning}
\begin{quote}
``You are a data analyst who reviews EEG data and helps interpret participant’s conditions.
Please review the power spectrum data of EEG data.
The data has an x-axis of time and a y-axis of frequency (0-40 Hz).
The data is collected from healthy participants who went through cognitive tasks.
We would like to classify the user state into 2 categories: focused or unfocused,   
confidence level (0-100\%), and explanations.
Please return your response in JSON format."
\end{quote}

\subsubsection*{In-context learning}
\begin{quote}
Please review the following 6 EEG data.
The data has an x axis of time and a y axis of amplitude.
The data is collected from healthy participants who went through sleep.
The first 5 images were collected during wake state (W), stage 1, stage 2, stage 3\&4, and REM sleep. Please provide 
1. a description of each image 
2. estimate the sleep stage of the 6th (last) image and return a number from 0 to 5 (0:wake, 1: stage 1, 2: stage 2, 3: stage 3 or 4, 4: REM, 5: unknown or movement)
3. confidence level (0-100\%) 
4. explanations 
in JSON format.
\end{quote}

\subsubsection*{Sleep quality detection}
\begin{quote}
``You are a data analyst helping a neurologist understand human sleep data.
Task: This is the data collected from users who fill out the Pittsburgh Sleep Quality Index questionnaire.
Based on this data, please give me the following information.
(1) if this person has good sleep quality or poor quality
(2) confident level (0-100\%)
(3) explanations for the decision.
(4) suggestions to improve sleep quality.”

\end{quote}

\subsubsection*{Sleep quality detection}
\begin{quote}
``Please review the following graph made from smartphone activity sensor data. 
The data has an x-axis of time (midnight to midnight) and a y-axis of date. For each hour, we computed the mean of activities where 0: Stationary, 1: Walking,  2:Running, 3:Unknown. Please describe
(1) estimated sleep quality (good or poor)
(2) confidence (0-100%)
(3) explanations
 (4) suggestions to improve sleep quality
in JSON format.”
\end{quote}

\subsubsection*{Suggestion to improve sleep/Guided Imagery}
\begin{quote}
 ``You are a sleep therapist.
 The participant is [participant conditions/profiles]. Reflecting on the participant's profile, please generate suggestions (a guided imagery script) to help the participant sleep better."
\end{quote}

\subsubsection*{Sleep Guided Imagery using EEG features}
\begin{quote}

``You are a sleep therapist.
Please generate a guided imagery script based on the following 30-s epoch EEG features,
alpha, beta, delta, theta, gamma power, alpha delta ratio, theta alpha ratio, delta theta ratio, mean, standard deviation, kurtosis, 90th percentile of amplitude.
Do not include actual numerical EEG features, brain waves, power spectrum, or different frequencies of waves in the script but include the state estimated from the features.
%Don't forget to adjust the script based on each 30s epoch but continue your script without discontinuity from the last one.
"
\end{quote}

\end{document}